\documentclass[superscriptaddress,groupedaddress,nofootnoteinbib,12pt]{article} 

\usepackage{jcappub}
\usepackage{bm}
\usepackage{verbatim}

\usepackage{epsf}
\usepackage{graphicx,epsfig}
\usepackage{amsfonts}
\usepackage{amssymb}
\usepackage{xcolor}
\usepackage{textcomp}

\definecolor{mine}{rgb}{0.2,0.1,0.7}
\definecolor{bb}{rgb}{0.3, 0.5, 1}
\definecolor{bg}{rgb}{0.1, 0.1, 0.5}

\def\a{\alpha}

\def\Q{{\cal Q}}

\def\kmi{k_{\rm min}}
\def\kma{k_{\rm max}}

\def\bkone{\mathbf k_1}
\def\bktwo{\mathbf k_2}
\def\bkthree{\mathbf k_3}
\def\d{{\rm d}}

\newcommand{\sdelta}[1]{\!\delta (\mathbf{#1})}
\def\p{\partial}

\def\C{\cal C}

\def\nn{\nonumber}

\newcommand{\dn}[2]{{\mathrm{d}^{{#1}}{{#2}}}}
\newcommand\bk{\boldsymbol{k}}

\newcommand{\bea}{\begin{eqnarray}}
\newcommand{\eea}{\end{eqnarray}}
\newcommand\be{\begin{equation}}
\newcommand\ee{\end{equation}}
\newcommand\beq{\begin{equation}}
\newcommand\eeq{\end{equation}}

\def\ba{\begin{eqnarray}}
\def\ea{\end{eqnarray}}

\newcommand{\refeq}[1]{(\ref{#1})}

\def\c{c_s}
\def\z{\zeta}
\def\log{{\rm log}}

\begin{document}

\title{A Modal Approach to the Numerical Calculation of Primordial non-Gaussianities}

\author[1]{Hiroyuki Funakoshi}
\author[2]{and S\'ebastien Renaux-Petel}
\affiliation[1]{Centre for Theoretical Cosmology,
Department of Applied Mathematics and Theoretical Physics,
University of Cambridge, Cambridge CB3 0WA, U.K. }
\affiliation[2]{Laboratoire de Physique Th\'eorique et Hautes Energies, Institut Lagrange de Paris, University Paris 6, 4 place Jussieu, Paris, France}
\vskip 4pt

\date{\today}


\abstract{We propose a new method to numerically calculate higher-order correlation functions of primordial fluctuations generated from any early-universe scenario. Our key-starting point is the realization that the tree-level In-In formalism is intrinsically separable. This enables us to use modal techniques to efficiently calculate and represent non-Gaussian shapes in a separable form well suited to data analysis. We prove the feasibility and the accuracy of our method by applying it to simple single-field inflationary models in which analytical results are available, and we perform non-trivial consistency checks like the verification of the single field consistency relation. We also point out that the $i \epsilon$ prescription is automatically taken into account in our method, preventing the need for ad-hoc tricks to implement it numerically.}

\maketitle


\section{Introduction}

The deviation from perfect Gaussian statistics of the primordial cosmological fluctuations enables us to discriminate amongst the candidate physical mechanisms that produced the seed primordial fluctuations (for recent reviews, see for instance \cite{Chen:2010xka,Byrnes:2010em,Wands:2010af,Barnaby:2010sq,Langlois:2011jt}). With the imminent arrival of Planck data together with those from large scale structures surveys, it is fundamental that theorists have firm predictions in wide classes of models to match against these increasingly precise data. However, most non-Gaussian shapes up to now have been determined under restrictive approximations, typically of the slow-varying type, to enable an analytical calculation. Interesting inflationary dynamics nonetheless exist which violate these assumptions and for which no theoretical prediction is yet available, for instance models with a strong background time-dependence or multifield models with non-trivial trajectories (see for example \cite{McAllister:2012am} for a recent illustration).

Procedures to numerically calculate the non-Gaussian properties of early-universe models
do exist \cite{Chen:2006xjb,Chen:2008wn,Martin:2011sn,Hazra:2012yn}, but they rely on direct $k$-space configuration by $k$-space configuration calculations requiring tricks to implement the $i \epsilon$ prescription necessary to project onto the interacting vacuum, like the introduction of a `damping' factor into the required time-integral to ensure early-time convergence.

In this paper, we propose an alternative method based on the separability of the tree-level In-In formalism and the modal decomposition techniques of Fergusson, Shellard and collaborators \cite{Fergusson:2006pr,Fergusson:2008ra,Fergusson:2009nv,Liguori:2010hx,Regan:2010cn,Fergusson:2010dm,Fergusson:2010ia,Fergusson:2010gn,Fergusson:2011sa,Regan:2011zq} (see also \cite{Meerburg:2010ks,Ribeiro:2011ax,Battefeld:2011ut}). It enables one to numerically compute tree-level higher-order correlation functions of cosmological observables from any early-universe scenario, with three main advantages:\\

 \textbullet \,\,\  no ad-hoc trick is required to implement the $i \epsilon$ prescription, which is automatically taken into account in our method.

 \textbullet \,\,\ the results of our procedure are smooth functions of their arguments, the $k_i$'s, and not values of the relevant correlation functions on a grid of points in Fourier space.
 
 \textbullet \,\,\ the resulting shapes are directly represented in a separable form, and hence can be efficiently constrained by data, for instance from the Planck satellite.\\

The outline of the paper is as follows: in section \ref{sec:method}, we present the key-idea behind our method, namely the use of the intrinsic separability of the tree-level In-In formalism, and details about its numerical implementation. In section \ref{sec:results}, we demonstrate its efficiency by calculating the tree-level bispectra generated by various operators in single-field test scenarios, like chaotic and Dirac-Born-Infeld inflation, and comparing them to known analytical results. We also perform non-trivial checks of our formalism, including the verification of the single field consistency relation between the scalar spectral index and the squeezed limit of the primordial bispectrum \cite{Maldacena:2002vr,Creminelli:2004yq}. We state our conclusions in section \ref{sec:conclusion}.

\section{Method}
\label{sec:method}

\subsection{Separability of the In-In formalism}

The central starting point of our method relies on the realization that the In-In formalism used to calculate primordial cosmological correlation functions \cite{Weinberg:2005vy} is intrinsically separable, at least at tree-level. 

The In-In (also known as Keldysh-Schwinger) formalism gives the following prescription for calculating the expectation value of any operator $Q(t)$ in the interacting vacuum of the theory:
\be
\langle Q(t) \rangle = \langle 0| \left[ \bar T \exp \left( i \int_{-\infty(1+i \epsilon)}^t H_I(t') dt' \right) \right]
~Q^I(t)~
\left[ T \exp \left( -i \int_{-\infty(1-i \epsilon)}^t H_I(t'') dt'' \right)\right] |0\rangle \,, \nonumber
\ee
where $ |0\rangle$ denotes the vacuum of the free theory, $T$ the time-ordering product, $H_I$ the interacting Hamiltonian, $Q^I(t)$ is in the interacting picture, and the $i \epsilon$ indicates that the time integration contour should be slightly rotated into the imaginary plane in order to project onto the interacting vacuum \cite{Maldacena:2002vr}.

Concentrating, for simplicity of presentation, on the bispectrum of the curvature perturbation $\zeta$ generated in single-field models of inflation, one is led to evaluate, at tree-level:
\be
\langle \zeta_{\bkone}(\tau) \zeta_{\bktwo}(\tau)  \z_{\bkthree}(\tau)  \rangle =   - i \int_{-\infty(1-i \epsilon)}^\tau \d \tau ' \, a(\tau') \langle 0| \z_{\bkone}(\tau) \z_{\bktwo}(\tau)  \z_{\bkthree}(\tau)  H_{(3)}(\tau') | 0 \rangle +{\rm c.c. }
\ee
where $\tau$ denotes the conformal time such that $\d \tau=\d t/a(t)$, $ H_{(3)}$ is the cubic interacting Hamiltonian, and all operators on the right-hand side are in the interacting picture (we omit the $I$ for brevity). In practice, the dimensionless quantity we will compute and which is ultimately the most useful for data analysis is the so-called shape $S(k_1,k_2,k_3)$, defined such that \cite{Babich:2004gb,Fergusson:2008ra}
\begin{equation}
\langle\zeta_{{\mathbf k_1}}\,\zeta_{{\mathbf k_2}}\, \zeta_{{\mathbf k_3}}\rangle \equiv (2\pi)^7 \, \sdelta{{\mathbf k_1}+{\mathbf k_2}+{\mathbf k_3}} \, \frac{S(k_1,k_2,k_3)}{(k_1 k_2 k_3)^2} A_s^2\,,
\label{def-S}
\end{equation}
where $A_s$ is a normalization factor related to the amplitude of scalar fluctuations. Illustrating this on the two cubic interactions $S_{(3)} \supset \int  \d {t} \, \dn{3}{x}\,a(t)\, g(t)\,  \dot{\z} (\p \z)^2$ and $S_{(3)} \supset \int  \d {t} \, \dn{3}{x}\,a(t)\,g(t) \, \z (\p \z)^2$, where $g(t)$ denotes a generic time-dependent coupling constant, we find respectively that 
 \begin{eqnarray}
 S_{\dot{\zeta} (\partial \zeta)^2}(k_1,k_2,k_3) &\propto& i \, \mathbf{k_2\cdot k_3} \,    \int_{-\infty(1-i\epsilon)}^{\tau} \mathrm{d}\tau' a(\tau') g(\tau') \Bigg( k_1^2  \zeta_{k_1}(\tau)  \zeta^{*'}_{k_1}(\tau') \Bigg) \cr
 &\times&  \Bigg( k_2^2  \zeta_{k_2}(\tau)  \zeta^{*}_{k_2}(\tau') \Bigg)   \Bigg( k_3^2  \zeta_{k_3}(\tau)  \zeta^{*}_{k_3}(\tau') \Bigg) +\, {\rm c.c.} +2\,{\rm perms.}
  \label{example1}
  \end{eqnarray}
and
 \begin{eqnarray}
 S_{\z (\partial \zeta)^2}(k_1,k_2,k_3) &\propto& i\, \mathbf{k_2\cdot k_3} \,   \int_{-\infty(1-i\epsilon)}^{\tau}\mathrm{d}\tau' \,a^2(\tau') \, g(\tau') \Bigg( k_1^2  \zeta_{k_1}(\tau)  \zeta^{*}_{k_1}(\tau') \Bigg) \cr
 & \times &  \Bigg(   k_2^2 \zeta_{k_2}(\tau)   \zeta_{k_2}^{*}(\tau') \Bigg) \Bigg( k_3^2 \zeta_{k_3}(\tau) \zeta_{k_3}^{*}(\tau') \Bigg) +\, {\rm c.c.}+2\,{\rm perms.}
  \label{example2}
  \end{eqnarray}
 where $\zeta^{'}_k$ denotes the time derivative of the mode function with respect to $\tau$, and $\mathbf{k_2\cdot k_3}=\frac12 \left(k_1^2-k_2^2-k_3^2 \right)$ using the triangle condition ${\mathbf k_1}+{\mathbf k_2}+{\mathbf k_3}=0$.

The key-point to notice is that, up to explicitly known $k$-dependent factors, like $\mathbf{k_2\cdot k_3}$ in Eqs.~\refeq{example1}-\refeq{example2}, the primordial shape generated by any cubic operator is explicitly given in a separable form, \textit{i.e.} as the time integral of a product of three functions of respectively $k_1,k_2,k_3$. This crucial fact therefore enables one to apply efficient separable techniques to determine a modal decomposition of the non-Gaussian profile $S$ generated by any cubic operator.

\subsection{Modal decomposition}
\label{subsec:modal}

Following the modal techniques of Fergusson, Shellard \textit{et al}, we wish to represent any primordial shape $S(k_1,k_2,k_3)$ (or similar quantities related to the relevant shape by multiplication by known $k$-dependent factors and/or symmetrizations) as
\begin{eqnarray}
S(k_1,k_2,k_3)=\sum_n\alpha_n\mathcal{Q}_n(k_1,k_2,k_3)
\label{decomposition}
\end{eqnarray}
where the $\mathcal{Q}_n$'s are suitable basis functions that are separable and orthonormal, so that the $\alpha_n$'s can be computed efficiently. For this purpose, we should halt and define both the space over which we want to expand the shapes and the scalar product with respect to which orthonormality is defined. Physically, the shapes $S$ are defined on a domain in Fourier space such that: (i) all $k_i$'s $\in [k_{{\rm min}},k_{{\rm max}}]$, where $k_{{\rm min}}$ and $k_{{\rm max}}$ stand for respectively the largest and smallest scales of interest (which depend on the particular probe we are interested in, for example CMB or large scale structures), (ii) the wavenumbers are such that the triangle condition ${\mathbf k_1}+{\mathbf k_2}+{\mathbf k_3}=0$ holds, namely
\begin{equation}
k_1 \leq k_2\,+\,k_3 \,\,\, {\rm for}\,\,\, k_1 \geq k_2,k_3, \,\,\,\,{\rm (and\,\, permutations)} \,.
\end{equation}
However, it is easy to realize that calculating the coefficients $\alpha_n$'s on such a tetrahedral domain would come at the cost of losing separability. Therefore, we consider instead the expressions \refeq{example1}-\refeq{example2} (and similarly for other vertices) as mathematically defining shapes over the cubic domain ${\cal C} \equiv [k_{{\rm min}},k_{{\rm max}}]^3$, irrespective of the triangle condition. Of course, for physical interpretations, the shapes we will thus construct should be restricted in the end to the tetrahedral domain, but the use of shapes defined on ${\cal C}$ enables us to use the intrinsic separability of the In-In formalism. Indeed, using the canonical scalar product on $\C$:
\begin{equation}
F(S,S') = \int_{\C} S(k_1,k_2,k_3) S'(k_1,k_2,k_3)\, \d k_1  \, \d k_2 \, \d k_3\,,
\label{canonical-sp}
\end{equation}
orthonormal and complete families of separable basis functions $\Q_n$ with the correct symmetry properties can be constructed out of products of Legendre polynomials. For instance, a completely symmetric function of $(k_1,k_2,k_3$), like the building block of Eq.~\refeq{example2}:
\begin{eqnarray}
\tilde{S}_{\z (\partial \zeta)^2}(k_1,k_2,k_3) & \equiv & \int_{-\infty(1-i\epsilon)}^{\tau}\mathrm{d}\tau' \,a^2(\tau') \, g(\tau') \Bigg( k_1^2  \zeta_{k_1}(\tau)  \zeta^{*}_{k_1}(\tau') \Bigg) \cr
 & \times &  \Bigg(   k_2^2 \zeta_{k_2}(\tau)   \zeta_{k_2}^{*}(\tau') \Bigg) \Bigg( k_3^2 \zeta_{k_3}(\tau) \zeta_{k_3}^{*}(\tau') \Bigg) \,,
\end{eqnarray}
can be represented on $\C$ as in Eq.~\refeq{decomposition}, using the basis functions
\bea
\Q_{n}&=&\frac{ \prod_{i=1}^{3}(2 n_i+1)^{1/2}}{(\kma-\kmi)^{3/2}} \Delta_{n_1 n_2 n_3} \nonumber \\
&\times & P_{(n_1}\left(\frac{2 k_1-(\kmi+\kma)}{\kma-\kmi}\right)P_{n_2}\left(\frac{2 k_2-(\kmi+\kma)}{\kma-\kmi}\right) P_{n_3)}\left(\frac{2 k_3-(\kmi+\kma)}{\kma-\kmi}\right) \nonumber
\eea
where $P_i$ denotes the $i$-th order Legendre polynomial, the index $n$ represents the symmetric triplet $\lbrace n_1,n_2,n_3 \rbrace$, and where $ \Delta_{n_1 n_2 n_3}$ equals $1,\sqrt{3},\sqrt{6}$ if all the $n_i$'s are equal, two of them are equal or all of them are different respectively. Using these orthonormal and \textit{separable} basis functions, 
together with the \textit{separable} representations of shapes as given by the In-In formalism, the modal decomposition coefficients $\alpha_n$'s in Eq.~\refeq{decomposition} can be computed as $\a_n= \int_{\C} \tilde{S} \Q_n$, where the triple integral over the cube $\C$ can be crucially decomposed as products of three one-dimensional integrals over $k_1,k_2$ and $k_3$ respectively:
\begin{eqnarray}
\a_n  &\propto &  \int_{-\infty(1-i\epsilon)}^{\tau}\mathrm{d}\tau' \,a^2(\tau') \, g(\tau') \Bigg( \int_{k_{\rm min}}^{k_{\rm max}}  P_{n_1}  k_1^2  \zeta_{k_1}(\tau)  \zeta^{*}_{k_1}(\tau') \,\d k_1\Bigg) \cr
 & \times & \Bigg( \int_{k_{\rm min}}^{k_{\rm max}}  P_{n_2}  k_2^2  \zeta_{k_2}(\tau)  \zeta^{*}_{k_2}(\tau') \,\d k_2\Bigg)  \Bigg( \int_{k_{\rm min}}^{k_{\rm max}}  P_{n_3}  k_3^2  \zeta_{k_3}(\tau)  \zeta^{*}_{k_3}(\tau') \,\d k_3\Bigg)\,,
\label{example-alpha-n}
\end{eqnarray}
and where the argument of the $P_n$'s are $(2 k_n-(\kmi+\kma))/(\kma-\kmi)$. Of course, a similar expansion holds for non-symmetric functions, like the building block of Eq.~\refeq{example1}
\begin{eqnarray}
\tilde{S}_{\dot{\zeta} (\partial \zeta)^2}(k_1,k_2,k_3) & \equiv & 
\int_{-\infty(1-i\epsilon)}^{\tau} \mathrm{d}\tau' a(\tau') g(\tau') \Bigg( k_1^2  \zeta_{k_1}(\tau)  \zeta^{*'}_{k_1}(\tau') \Bigg) \cr
 &\times&  \Bigg( k_2^2  \zeta_{k_2}(\tau)  \zeta^{*}_{k_2}(\tau') \Bigg)   \Bigg( k_3^2  \zeta_{k_3}(\tau)  \zeta^{*}_{k_3}(\tau') \Bigg) \,,
\end{eqnarray}
which is symmetric only in $(k_2,k_3)$. The only difference in such cases is that the basis polynomials should only be symmetric in $(k_2,k_3)$. Following these methods, the shape generated by any cubic operator can be represented by a modal decomposition whose coefficients can easily be calculated numerically. Finally, let us note that, although we have illustrated our reasoning by considering the curvature perturbation $\z$ only, it should be clear that similar techniques based on separability hold in theories with several degrees of freedom, like in multiple field inflationary models.

\subsection{Numerical considerations}
\label{numerics}  

It turns out that most of the difficulties we encounter to numerically evaluate the Fourier space integrals needed to determine the modal decomposition coefficients can be understood by using the crudest approximation for the mode functions, namely
\be
\z_{k}(\tau) \propto \frac{1}{\sqrt{k^3 }}(1+i k \c \tau)e^{-i k \c  \tau}\,, \qquad \z_{k}'(\tau) \propto  \frac{1}{\sqrt{k^3 }} k^2 \c^2 \tau  e^{-i k \c \tau}\,,
\label{mode-functions}
\ee
up to some slowly varying multiplicative factor function of both $k$ and $\tau$, and where $c_s$ is the slowly-varying speed of sound of fluctuations. Required integrals of the type 
\begin{equation}
 \int_{k_{\rm min}}^{k_{\rm max}}  P_{i}\left(\frac{2 k-(\kmi+\kma)}{\kma-\kmi}\right) k^2  \zeta_{k}(\tau)  \zeta^{*}_{k}(\tau') \,\d k
 \label{typical-integral}
\end{equation}
in \refeq{example-alpha-n} can be seem as sum of terms of the form
\begin{equation}
 \int_{k_{\rm min}}^{k_{\rm max}}  k^{2+n}  \zeta_{k}(\tau)  \zeta^{*}_{k}(\tau') \,\d k
 \label{k-integration}
\end{equation}
for $n \geq 0$. An obvious difficulty in numerically evaluating these integrals is the rapid oscillations coming from the phase factor $e^{- i k \c \tau}$ in the mode functions. However, this can be circumvented by numerically solving the linear mode function equation not for $\z_{k}(\tau)$ itself, but rather for $\tilde{\z}_{k}(\tau)=\z_{k}(\tau) e^{i k \c  \tau}$, which displays no rapid oscillations. This results in the integrand in Eq.~\refeq{k-integration} being put in the form of the slowly varying function $k^{2+n} \tilde{\z}_{k}(\tau) \tilde{\z}_{k}^*(\tau')$ multiplied by the explicitly known phase factor $ e^{i k (\c(\tau')  \tau'-\c(\tau)  \tau)}$, a type of integrand for which known efficient numerical routines can be used. In practice of course, the above slowly-varying function of $k$ is only known as a result of evaluating the mode functions at sample $k$-points and constructing an interpolating function out of them. This results in one last difficulty: for $n=0$, $k^{2+n} \z_{k}(\tau) \z_{k}^*(\tau')$ contain terms behaving like $1/k$ which do not admit a rapidly convergent polynomial representation. We overcome this by simply considering the well behaved function $ k^3 \zeta_{k}(\tau)  \zeta^{*}_{k}(\tau')$ (instead of $k^2  \zeta_{k}(\tau)  \zeta^{*}_{k}(\tau')$ in Eq.~\refeq{typical-integral}), and dividing by the appropriate factor of $k$ at the end of the calculations. On the other hand, note that no such trick is required for evaluating the integrals of the form $\int_{k_{\rm min}}^{k_{\rm max}}  P_{i} \, k^2  \zeta_{k}(\tau)  \zeta^{*'}_{k}(\tau') \,\d k$ because of the additional $k^2$ factor in $\z_{k}'(\tau)$ in Eq.~\refeq{mode-functions}. 
 
 Eventually, we observed that a slightly different expansion leads to improved accuracy, which consists in considering the shapes as function of the $y_i=\log(k_i) (i=1,2,3)$, rather than the $k_i$'s themselves, and therefore in decomposing them similarly as described above, but with the scalar product
\begin{equation}
F_{\rm \log}(S,S') = \int_{[\log(k_{{\rm min}}),\log(k_{{\rm max}})]^3} S(y_1,y_2,y_3) S'(y_1,y_2,y _3) \, \d y_1 \, \d y_2 \, \d y_3\,.
\end{equation} 
In the following, we denote results obtained by this method as resulting from the $\log$-method.
  
\subsection{Projecting onto the interacting vacuum and the $i \epsilon$ prescription}
\label{vacuum}

When directly computing the value of a non-Gaussian shape in a specific momentum configuration, using {\textit e.g.} Eq.~\refeq{example2}, the $i \epsilon$ prescription is needed to project on to the interacting vacuum. However, the quantities we are calculating in our approach are different: these are the coefficients that enter into the shape's modal decomposition \refeq{decomposition}, like $\alpha_n$ in Eq.~\refeq{example-alpha-n}. Here, the time-integrand comes as the result of integrating the mode functions in Fourier-space, which averages out the rapid oscillations of the mode functions deep inside the horizon, with one important consequence: evaluating Eq.~\refeq{example-alpha-n} for $\epsilon>0$ and taking the limit $\epsilon\rightarrow0$, following the $i \epsilon$ prescription, is equivalent to just evaluating it at $\epsilon=0$ straight away. Or, to put it a different way: the process of first integrating over Fourier space before performing the required time-integral is equivalent to the $i \epsilon$ prescription for us.

Hence, while integrals like the one in Eq.~\refeq{example2} do not converge, as their lower bound $\tau_{\rm ini}$ goes to $- \infty$, without an appropriate rotation of the time integration contour into the imaginary plane, this is not the case in our method. Actually, by using the asymptotic form of the mode functions \refeq{mode-functions} deep inside the horizon --- which only results from our choice of the Bunch-Davies vacuum --- and the behavior of the scale factor $a(\tau) \propto 1/\tau$, one can easily determine the early-time behavior of the required time-integrands, like the one in Eq.~\refeq{example-alpha-n}. One thus finds that the coefficients of the modal decomposition converge, as $\tau_{\rm ini}$ goes to $- \infty$, like $\int^{\infty} \frac{\d x}{x^2}e^{ix}$ for the operators in $\int \d t \, a(t) \, \zeta^3$ and $\int \d t \, a^3(t) \,\zeta \dot{\zeta}^2$ (whatever their explicit gradient structures), and like $\int^{\infty} \frac{\d x}{x}e^{ix}$ for the operators in $\int \d t \, a^3(t) \, \dot{\zeta}^3$ and $\int \d t \, a(t) \, \dot{\zeta} \zeta^2$. In practice, these behaviors, which we have verified in our numerical calculations, imply that requisite time-integrations can be performed only starting from a few e-folds before the largest scale of interest crosses the horizon.

\section{Results}
\label{sec:results}

In this section, we demonstrate that the method described above works as intended. We first describe the details of our set-up, in particular the inflationary backgrounds and the cubic operators we have considered, before showing explicit results in various situations.

\subsection{Set-up}

\noindent  \textit{Test scenarios}.---We consider two kinds of inflationary scenarios: the one of a canonical scalar field with an $m^2 \phi^2$ potential, which we refer to as the chaotic scenario, and the one corresponding to a Dirac-Born-Infeld (DBI) action in an AdS geometry \cite{Silverstein:2003hf}. In more detail, the two corresponding actions read (we omit the Einstein-Hilbert action which is common to the two set-up, and use units in which $M_p=1$)
\begin{eqnarray}
 \hspace{-2.0em} S_{\rm chaotic}&=& \int \d^4 x \sqrt{-g} \left( -\frac{1}{2} \partial_{\mu} \phi \partial^{\mu} \phi- \frac12 m^2 \phi^2 \right) \,,\\
 \hspace{-2.0em} S_{\rm DBI}&=& \int \d^4 x \sqrt{-g} \left( -\frac{1}{f(\phi)} \left(\sqrt{1+f(\phi)  \partial_{\mu} \phi \partial^{\mu} \phi   }-1 \right)- \frac12 m^2 \phi^2  \right) \,\,{\rm with}\,\, f(\phi)=\frac{\lambda}{\phi^4}. \label{SDBI}
\end{eqnarray}
Our aim is not to consider inflationary scenarios that match precisely observational data, for example in terms of the amplitude and spectral index of the scalar fluctuations they generate or their level of non-Gaussianities, but rather to consider backgrounds in which all the slowly-varying approximations that are usually made in analytical treatments are satisfied, so that our numerical procedures can be tested against reliable theoretical predictions.

Unless otherwise stated, we take $m=10^{-5}$ in both scenarios and $\lambda=10^{16}$. We also use initial conditions such that background attractor solutions are reached rapidly, namely $\dot{\phi}_i=-\sqrt{\frac32}m$ and $\phi_i=10^3$ in the chaotic case and $\dot{\phi}_i=-\left(\frac{3}{2 m^2}+\frac{\lambda}{\phi_i^4}  \right)^{-1/2}$ and $\phi_i=2$ in the DBI one. These parameters result in almost de-Sitter inflationary backgrounds in which the important background quantities $\epsilon \equiv -\dot{H}/H^2$ and $c_s^2=1-f(\phi) \dot{\phi}^2$ take almost constant very small values during more than $60$ efolds of expansion: $\epsilon_{\rm chaotic} \sim 10^{-6}$, $\epsilon_{\rm DBI} \sim 10^{-3}$ and $c_s \sim 10^{-3}$.\\

As for the linear fluctuations about these inflationary backgrounds, the curvature perturbation obeys the simple second-order equation
\begin{eqnarray}
\zeta_k''+2 \frac{z'}{z} \zeta_k'+c_s^2 k^2 \zeta_k=0\,, \quad z=\frac{a \sqrt{2 \epsilon}}{c_s}\,,
\label{zeta-equation}
\end{eqnarray}
where it is understood that $c_s^2 \equiv 1$ in the chaotic case, and we impose initial conditions such that the Bunch-Davies behavior holds inside the horizon:
\begin{eqnarray}
\zeta_k=i H\sqrt{\frac{c_s}{4 \epsilon k}} \tau e^{-i k c_s \tau} \quad {\rm for} \quad -k c_s \tau \gg 1\,.
\end{eqnarray}
Eventually, we consider a range of scales such that $\frac{k_{\rm min} c_{si}}{a_i H_i}=10^3$ --- all studied scales are therefore well inside the horizon initially\footnote{This is in agreement with our discussion about early-time convergence in subsection \refeq{vacuum}.} --- and with $k_{\rm max}/k_{\rm min}$ equals to $10^2$ (similar results are obtained with $k_{\rm max}/k_{\rm min}=10^3$).\\

\noindent  \textit{Cubic operators}.---Using the background and the linear fluctuations just described, we would like to test our method for numerically calculating the non-Gaussian shapes generated by various cubic operators. As we have explained in section \refeq{sec:method}, we factor out any explicitly known $k$-dependence coming from gradient terms in cubic operators. We therefore have to apply our method only to four kinds of cubic combinations of $\zeta$, namely $\zeta^3, \zeta^2 \dot{\zeta}, \zeta \dot{\zeta}^2$ and $\dot{\zeta}^3$, from which one can then reconstruct the shapes of any specific operator. In the context of the most general second-order scalar-tensor theory, also known as Horndeski theory \cite{Horndeski,Deffayet:2011gz,Kobayashi:2011nu}, it has been shown in Ref.~\cite{RenauxPetel:2011sb} that only $5$ independent cubic interactions can possibly be generated, \textit{i.e.} that the third-order action can effectively be put in the form
   \begin{eqnarray}
       \hspace{-2.0em}   S_{(3)} & = &    \int {\rm d}t \,\dn{3}{x}\, a^3   \biggl\{g_1\dot \zeta ^{3}+g_2 \z \dot \zeta^{2}+g_{3}\zeta\frac{(\p \z)^2}{a^2}     
           + g_4 \dot \zeta \partial_i \zeta \partial^i \left( \p^{-2} \dot \z \right) + g_{5}\partial^{2}\zeta  \left(\partial_i \p^{-2} \dot \z  \right)^{2}         \biggr\},\label{S3}
        \end{eqnarray}
        where the $g_i$'s are time-dependent coupling constants, and with other interactions, like for example $\int \d t \, \d^3 x \,a\,\dot{\zeta} (\partial \z)^2$, being redundant, in the sense that they can be expressed in terms of these $5$ by using the linear equation of motion \refeq{zeta-equation}. This fact provides the opportunity of interesting consistency checks, and we have therefore additionally considered this latter operator.\\

\noindent  \textit{Ordering of the basis functions}.---We have described in subsection \refeq{subsec:modal} our choice for the basis functions ${\cal Q}_n$ we are using to expand any shape in a separable form. Of course, in any practical application, we should truncate the infinite expansion \refeq{decomposition} and use a specific ordering for the ${\cal Q}_n$'s. For that purpose, we use the `slicing' ordering described in Ref.~\cite{Fergusson:2009nv} and decompose functions with polynomials (of $k_i$ or $\log(k_i)$ depending on the chosen method) up to a specific total order $N_{\rm max}$. For symmetric functions for example, the index $n$ thus relates to the completely symmetric triplets $\lbrace n_1,n_2,n_3 \rbrace$ (with a specific choice of sub-ordering):
\begin{align}
\nn &\underline{0  \rightarrow 000}   \qquad \nn 4  \rightarrow 111  \qquad~\,  \nn 8  \rightarrow 022\qquad  \nn 12  \rightarrow 113 \\
\nn  & \underline{1\rightarrow 001}   \qquad  \nn 5  \rightarrow 012 \qquad~\,  \nn 9  \rightarrow 013  \qquad \nn 13  \rightarrow 023 \\
& 2 \rightarrow 011  \qquad \underline{6  \rightarrow 003}\qquad \underline{10  \rightarrow 004}  \qquad  14  \rightarrow 014 \\   \nn
 &\underline{ 3 \rightarrow 002}  \qquad  7  \rightarrow 112 \qquad   11  \rightarrow 122 \qquad   \underline{15  \rightarrow 005} ~\cdots\,,
\end{align}
where the transitions between total polynomial order is underlined.

\subsection{Numerical results}

\noindent  \textit{Bare operators}.---To show the effectiveness of our method, we applied it to the six operators listed above, with constant $g_i$'s in Eq.~\refeq{S3}, for the two different test scenarios and using $N_{\rm max}=6$ in the modal decomposition\footnote{We stress that this is only methodological: for instance, not all $6$ operators are present in the chaotic or DBI third-order action.}. We then evaluated the correlations --- as measured by $F(S,S')/\left(\sqrt{F(S,S)F(S',S')}\right)$ with the scalar product $F$ \refeq{canonical-sp} restricted to the tetrahedral domain --- between the corresponding theoretical and numerically calculated shapes. Our results are listed in table \refeq{Cosines}. The various cosines are impressively close to $1$, demonstrating the ability of our method to reconstruct non-Gaussian shapes. Additionally, its rapid convergence is displayed in table \refeq{Convergence}, where we indicate the correlations between theoretical and numerically calculated shapes for increasing $N_{{\rm max}}$. We give the results for two representative operators of the equilateral and local type respectively, namely ${\dot \zeta}^3$ and $\zeta (\p \z)^2$, and for the test chaotic inflationary background (similar results are obtained in the DBI case). Eventually, let us note that it takes about $1.5$ minutes on a laptop to calculate all possible bispectrum shapes (remember that they can be all deduced from the $4$ cubic combinations of $\z$ and $\dot{\z}$) with $N_{{\rm max}}=2$, and $4$ minutes with $N_{{\rm max}}=6$. While this is sufficient for our purpose, we expect that a significant speed-up can be obtained after optimization of our code.

\begin{table}
\begin{center}
\begin{tabular}{c|c|c}
operator & correlation ($m^2\phi^2$) & correlation (DBI)\\
\hline
$\dot{\zeta}^3$ & $0.9992$ & $0.9994$\\
$\dot{\zeta}(\partial\zeta)^2$ & $0.99997$ & $0.99995$\\
$\zeta \dot{\zeta}^2$ & $0.999994$ & $0.999990$\\
$\zeta(\partial\zeta)^2$ & $0.999998$ & $0.999995$\\
$\dot{\zeta}\partial_i\zeta\partial^i(\partial^{-2}\dot{\zeta})$ & $0.99998$ & $0.99997$\\
$\partial^2\zeta(\partial_i\partial^{-2}\dot{\zeta})^2$ & $0.999990$ & $0.99998$\\
\end{tabular}
\end{center}
\caption{Correlations between theoretical and numerically calculated shapes generated by various (bare) operators on our two test background inflationary scenarios.}
\label{Cosines}
\end{table}

\begin{table}[h!]
\begin{center}
\begin{tabular}{c|c|c|c|c|c|c|c}
operator & $N_{{\rm max}}=0$ & $N_{{\rm max}}=1$ & $N_{{\rm max}}=2$ & $N_{{\rm max}}=3$ & $N_{{\rm max}}=4$ & $N_{{\rm max}}=5$ & $N_{{\rm max}}=6$  \\
\hline
$\dot{\zeta}^3$ & $0.98$ & $0.97$ & $0.994$ & $0.998$ & $0.9990$ & $0.9993$ & $0.9994$\\
$\zeta (\partial\zeta)^2$ & $0.90$ & $0.997$ & $0.9997$ & $0.99994$ & $0.999990$ & $0.999996$ & $0.999998$\\
\end{tabular}
\end{center}
\caption{Correlations between theoretical and numerically calculated shapes generated by the two bare operators ${\dot \zeta}^3$ and $\zeta (\p \z)^2$ on the test chaotic background, for varying $N_{{\rm max}}$.}
\label{Convergence}
\end{table}

\noindent  \textit{Two non-trivial checks}.---In single-field inflation, there exists a well known consistency relation between the amplitude of the primordial bispectrum in the squeezed limit and the scalar spectral index \cite{Maldacena:2002vr,Creminelli:2004yq}, namely that
\be
\lim_{k_3 \ll k_1 \simeq k_2} \langle \zeta_{\bkone} \zeta_{\bktwo}\zeta_{\bkthree} \rangle = -(2 \pi)^3  \delta^{(3)} (\sum_i \bk_i)(n_s(k_1)-1)P_{\zeta}(k_1)P_{\zeta}(k_3) 
\label{consistency}
\ee
where $n_s(k)$ is defined as
\be
n_s(k)-1\equiv  \frac{ d \, {\rm ln }\left[ k^3P_{\zeta}(k)\right]}{ d\, {\rm ln} k}
\ee
with
\be
\langle \zeta_{\bkone} \zeta_{\bktwo}\rangle \equiv (2 \pi)^3  \delta^{(3)} (\bkone + \bktwo)P_{\zeta}(k_1) \,.
\ee
To demonstrate the accuracy of our method, we calculated the bispectrum generated in various chaotic inflationary scenarios, which, at leading-order in a slow-varying approximation, simply corresponds to setting $g_2(t)=\epsilon (\epsilon-\eta)$ and $g_3(t)=\epsilon (\epsilon+\eta)$ in Eq.~\refeq{S3}, where $\eta \equiv \dot{\epsilon}/(H \epsilon)$. We then evaluated the reconstructed bispectra on squeezed triangles such that $k_3= 0.01 k_1=0.01k_2$ and we numerically calculated the corresponding $n_s(k_1)$. We found that in all cases the single field consistency relation was verified to better than the \% level. For instance, for $\phi_i=16$, we find that $n_s(k_{{\rm max}})-1=-0.0379$, to be compared with $-0.0377$ deduced from the numerically calculated bispectrum, so a $0.5 \%$ discrepancy. \\

As we mentioned above, another non-trivial check of our numerical method involves the redundancy of some cubic operators \cite{RenauxPetel:2011sb}. At leading order in the slow-varying approximation, the DBI third-order action reads \cite{Chen:2006nt}
  \begin{eqnarray}
S_{(3),{\rm DBI}} \simeq  \int  \d{t} \, \dn{3}{x} \ a^3  \epsilon \left(\frac{1}{c_s^2}-1 \right) \left[-\frac{3}{c_s^2} \z \dot \z^2+ \frac{\z (\partial \z)^2}{a^2} \right].
\label{S3-DBI-1}
\end{eqnarray}
Surprisingly enough at first sight, the two cubic operators present in this action generate non-Gaussian shapes that are strongly correlated with the local ansatz \cite{Komatsu:2001rj}, whereas the DBI scenario is known to generate a bispectrum of the equilateral type \cite{Alishahiha:2004eh}. A way of understanding this is to note that, upon using the linear mode function equation \refeq{zeta-equation}, an equivalent DBI third-order action reads, at leading-order in the slow-varying approximation\footnote{Note that the first form \refeq{S3-DBI-1} of $S_{(3)}$ naturally results from calculating it in the uniform inflaton gauge, while the form \refeq{S3-DBI-2} naturally appears when working in the spatially flat gauge.} \cite{RenauxPetel:2011sb}:
\begin{eqnarray}
S_{(3),{\rm DBI}} \simeq  \int  \d{t} \, \dn{3}{x} \ a^3 \frac{\epsilon}{H} \left(\frac{1}{c_s^2}-1 \right) \left[-\frac{1}{c_s^2} \dot  \z^3+ \frac{\dot \z (\partial \z)^2}{a^2} \right].
\label{S3-DBI-2}
\end{eqnarray}
This latter form of the third-order action makes it manifest that the total DBI bispectrum is of equilateral type, as this is the case for the individual bispectra generated by the operators in $\dot  \z^3$ and $\dot \z (\partial \z)^2$. 

\begin{figure}
\begin{center}
\includegraphics[width=1.00\linewidth]{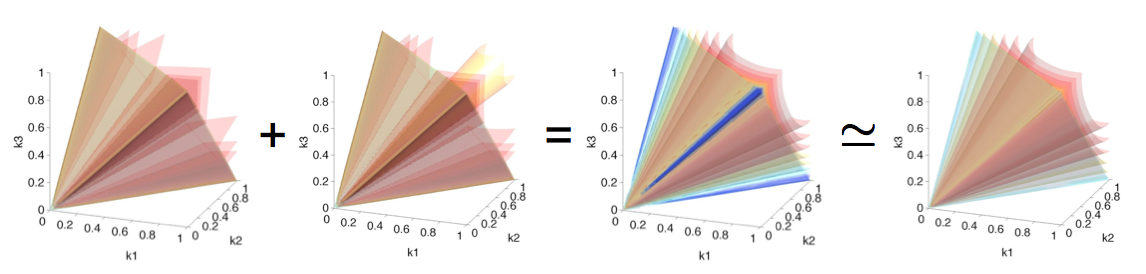}  
\caption{A graphical representation of a non-trivial check of our method. From left to right: the shapes generated in our test DBI scenario by, respectively, the $\z \dot \z^2$ and $\z (\partial \z)^2$ operators in Eq.~\refeq{S3-DBI-1}, the resulting total shape, and the total shape calculated from Eq.~\refeq{S3-DBI-2}. Each shape are represented through several density contours. The third one shows that the cancellations of the local-type behaviors of the first two shapes is not perfect in the squeezed limits (in blue). The overlap with the fourth shape is nonetheless remarkable, with a correlation of $0.993$ between the two.}
\label{fig-cancellations}
\end{center}
\end{figure}

The fact that the local-type behaviors of the two operators $\z \dot \z^2$ and $\z (\partial \z)^2$ should conspire to cancel in \refeq{S3-DBI-1}, resulting in a total equilateral-type shape, provides a non-trivial test of our method. We have verified it, obtaining a correlation of $0.993$ between the bispectra calculated using Eq.~\refeq{S3-DBI-1} and Eq.~\refeq{S3-DBI-2} in our test DBI scenario. This is summarized graphically in Fig.~\refeq{fig-cancellations}, where shapes are represented through several density contours.\\

\noindent  \textit{Other examples}.---For illustrative purposes, we display below examples of theoretical bispectrum shapes (left) and their numerically calculated counterparts (right) in our test DBI scenario. We used the $\log$-method (\textit{c.f.} subsection \refeq{numerics}) with $N_{\rm max}=8$, together with the coupling constants of Eqs.~\refeq{S3-DBI-1}-\refeq{S3-DBI-2}. The agreement is impressive.\\

 \hspace{-1.0em}  $\dot{\zeta}^3$ \hspace{+1.0em}  \noindent\begin{tabular}{@{\hspace{0.0em}}c@{\hspace{1.5em}}c@{\hspace{0.0em}}}
Theoretical bispectra & Numerically calculated bispectra\\ 
 \includegraphics[width=0.39\linewidth]{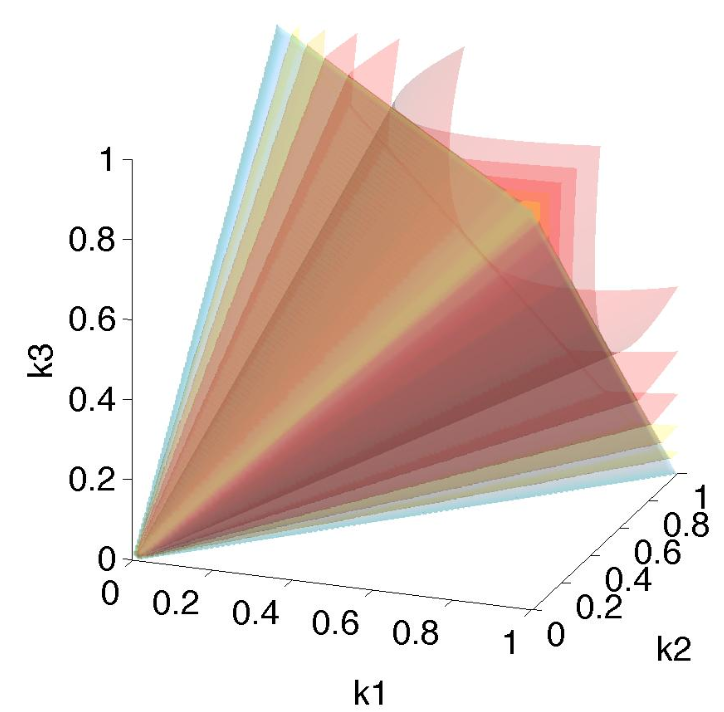} &
\includegraphics[width=0.39\linewidth]{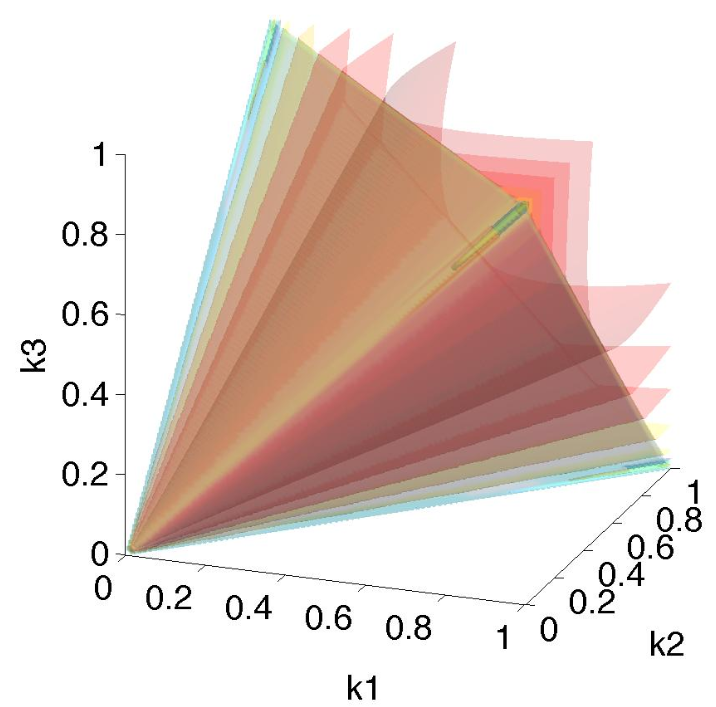} \\
\end{tabular}
 
 \hspace{-1.5em}  $\dot{\zeta} (\partial \zeta)^2$  \noindent\begin{tabular}{@{\hspace{0.0em}}c@{\hspace{1.5em}}c@{\hspace{0.0em}}}
 \includegraphics[width=0.39\linewidth]{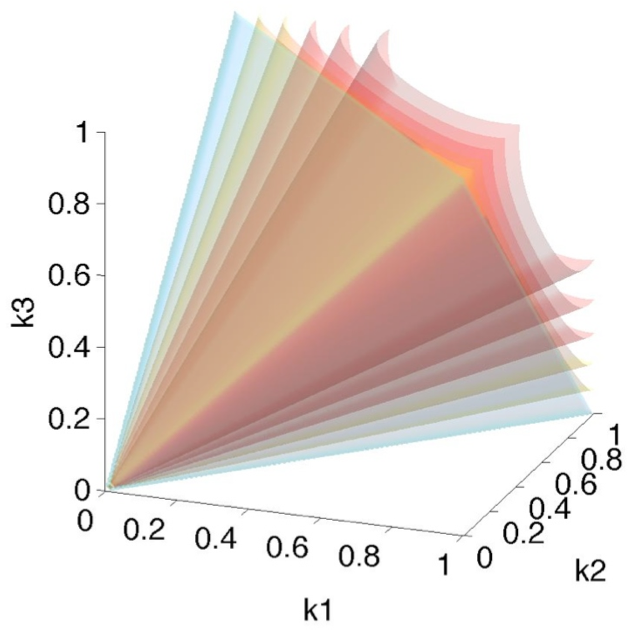} &
\includegraphics[width=0.39\linewidth]{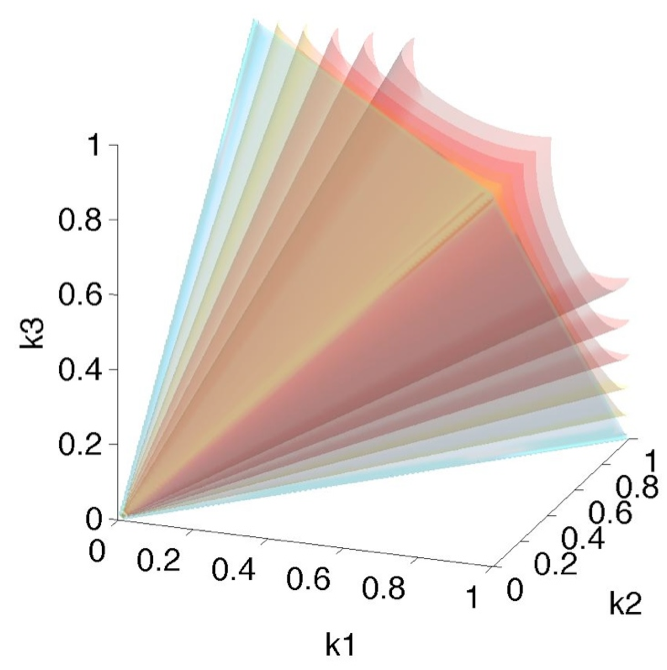} \\
\end{tabular}

 \hspace{-1.5em} $\zeta {\dot \zeta}^2$  \hspace{+1.0em}    \noindent\begin{tabular}{@{\hspace{0.0em}}c@{\hspace{1.5em}}c@{\hspace{0.0em}}}
\includegraphics[width=0.39\linewidth]{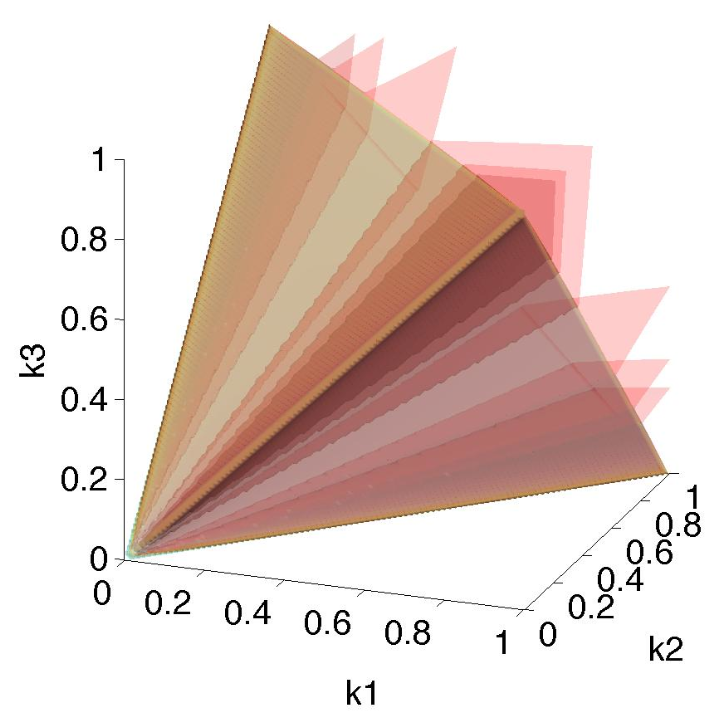} &
\includegraphics[width=0.39\linewidth]{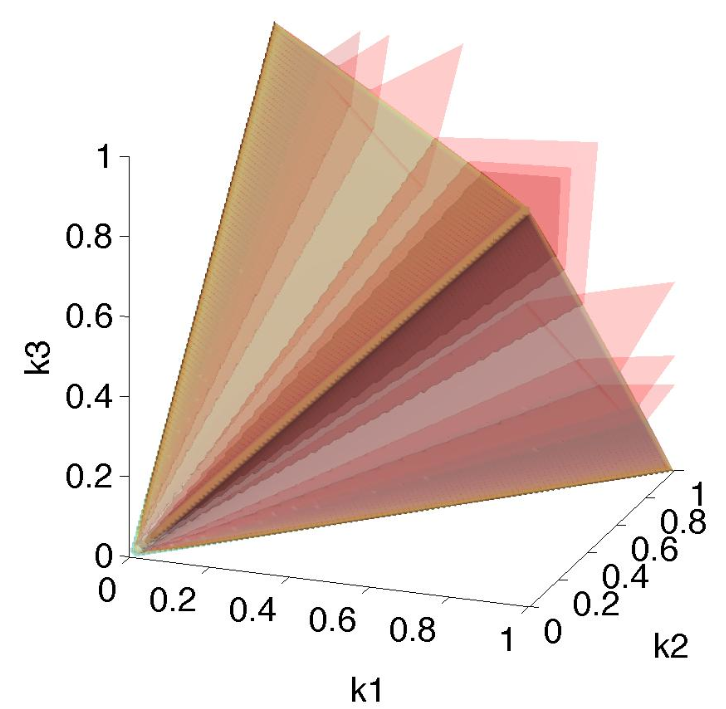} \\
\end{tabular}   
 
 \hspace{-1.5em} $\zeta (\partial \zeta)^2$    \noindent\begin{tabular}{@{\hspace{0.0em}}c@{\hspace{1.5em}}c@{\hspace{0.0em}}}
\includegraphics[width=0.39\linewidth]{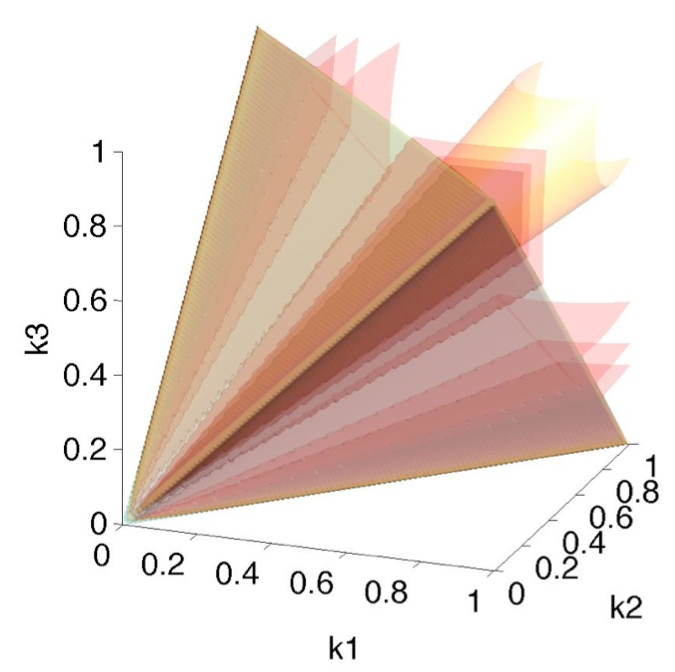} &
\includegraphics[width=0.39\linewidth]{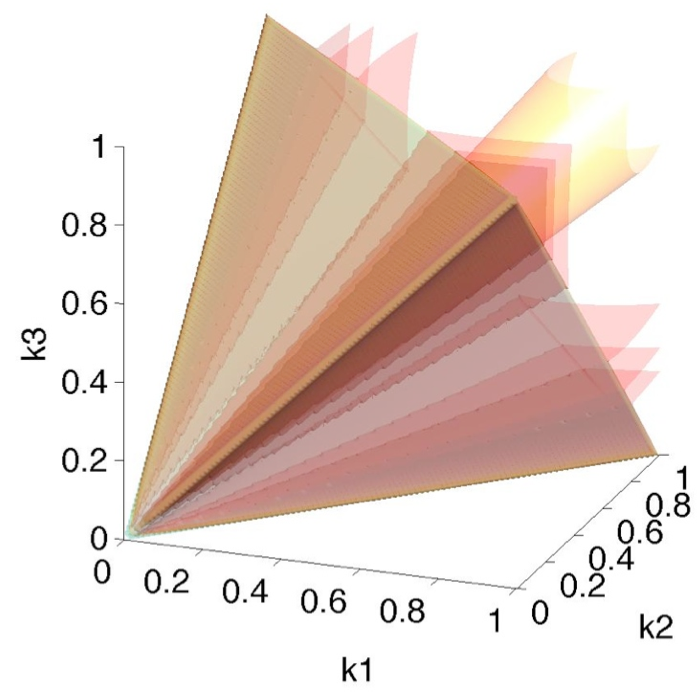} \\
\end{tabular}  

\begin{figure}[h!]
\begin{center}
\includegraphics[width=0.90\linewidth]{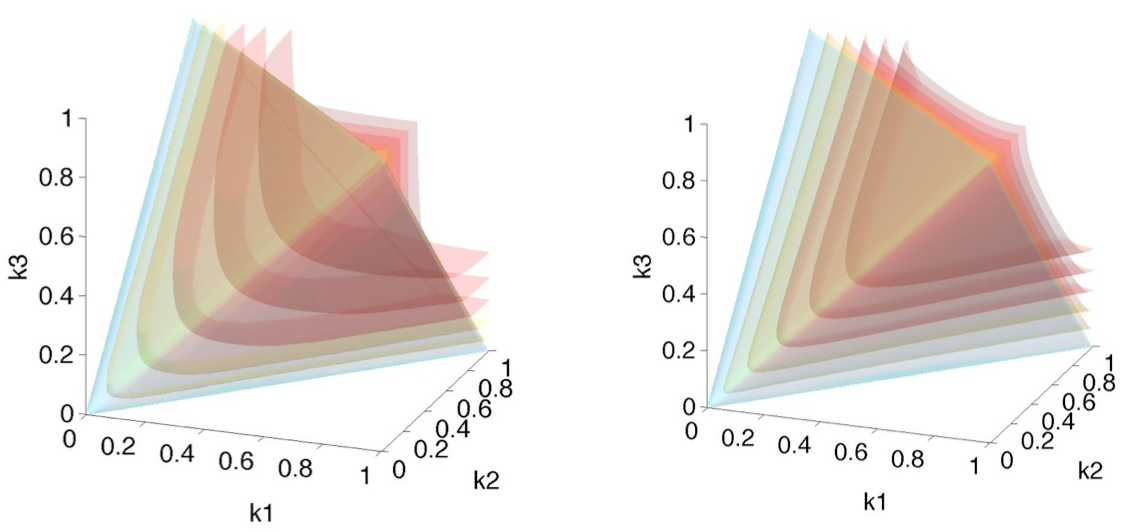}  
\caption{Shapes generated by the two operators $\dot{\zeta}^3$ (left) and $\dot{\zeta} (\p \z)^2$ (right) in the third-order action Eq.~\refeq{S3-DBI-2}, on the DBI background with $\lambda=10^{12}$ (see the main text). The rapidly varying background leads to strongly scale-dependent equilateral-type bispectra.}
\label{S1-2-rapidly}
\end{center}
\end{figure}

Note that because the slow-varying approximation is very accurate on our test DBI background, the numerically calculated shapes displayed above are almost exactly scale invariant. For comparison, we show in Fig.~\refeq{S1-2-rapidly} the shapes generated by the two operators $\dot{\zeta}^3$ and $\dot{\zeta} (\p \z)^2$ in the third-order action Eq.~\refeq{S3-DBI-2} under the same conditions as previously, but with $\lambda=10^{12}$ in the action \refeq{SDBI}. This modification results in a much more rapidly varying background, and hence to strongly scale-dependent equilateral non-Gaussianities, whose amplitude in the equilateral limit grows by a factor of about $30$ between the largest and the smallest scales. This exemplifies the effectiveness of our approach for studying primordial non-Gaussianities generated from non-trivial inflationary dynamics.

\section{Conclusions}
\label{sec:conclusion}

In this paper we have proposed a new method to numerically calculate higher-order correlation functions of primordial fluctuations generated from early-universe scenarios. It is based on the realization that the tree-level In-In formalism is intrinsically separable, which enables one to use modal techniques to efficiently represent any non-Gaussian profile. We have proved the feasibility and the accuracy of our method by applying it to simple single-field inflationary scenarios in which analytical results are available, finding impressive agreements and rapid convergence with respect to the number of modes used in the decomposition. We also performed non-trivial consistency checks, including the verification of the single field consistency relation, and the cancellations of two local-type behaviors to obtain an equilateral-type shape, as relevant in the DBI inflationary model. Additionally, we found our code to be quick, although a significant speed-up can be expected after proper optimization.\\

To us, the main advantages of our method are that: (i) the $i \epsilon$ prescription is automatically taken into account, preventing the need for ad-hoc tricks to implement it. (ii) We are able to calculate correlation functions at once over the whole domain of interest, and not only for specific $k$-space configuration. (iii) Non-Gaussian shapes are directly expressed in a separable form, and hence are ready to be efficiently constrained by data.\\

While we have mainly applied our method to simple early-universe models, thus providing a proof of its concept, its usefulness ultimately lies in the fact that it provides an efficient way of determining the non-Gaussian properties of cosmological fluctuations in non-trivial models in which analytical techniques are known to be insufficient. The result of such investigations will be presented elsewhere \cite{preparation1,preparation2}.

\medskip

\begin{acknowledgments}

We would like to thank Paul Shellard for helping initiating this project. We are also grateful to Xingang Chen, James Fergusson, Eugene Lim, J\'er\^ome Novak and Marcel Schmittfull for useful discussions. James Fergusson is particularly thanked for helping producing the 3D figures, and Cyril Pitrou for comments on a draft of this work. SRP would also like to thank the participants of the workshop `Critical Tests of Inflation Using Non-Gaussianity' for their useful comments. SRP was supported by the STFC grant ST/F002998/1 and the Centre for Theoretical Cosmology when most part of this work was carried out. This work was supported by French state funds managed by the ANR within the Investissements d'Avenir programme under reference ANR-11-IDEX-0004-02.

\end{acknowledgments}

\end{document}